\documentclass[pra,twocolumn,aps,showpacs,superscriptaddress,10pt]{revtex4-1}

\usepackage[utf8]{inputenc}
\usepackage{graphicx}  
\graphicspath{{./Figures/}}
\usepackage{dcolumn}  
\usepackage{amssymb, amsmath}
\usepackage{hyperref}
\usepackage{lineno} 
\usepackage{braket}
\usepackage{multirow}
\usepackage{bbold}
\usepackage{color}

\begin{document}
\title{Heading errors in all-optical alkali-vapor magnetometers in geomagnetic fields}
\author{W.\ Lee}\email{wonjael@princeton.edu}
\author{V.\ G.\ Lucivero}\altaffiliation[Current affiliation: ]{ICFO-Institut de Ciencies Fotoniques, The Barcelona Institute of Science and Technology, 08860 Castelldefels (Barcelona), Spain}
\author{M.\ V.\ Romalis}
\affiliation{Department of Physics, Princeton University, Princeton, New Jersey 08540, USA}
\author{M.\ E.\ Limes}
\author{E.\ L.\ Foley} 
\author{T.\ W.\ Kornack}
\affiliation{Twinleaf LLC, Princeton, New Jersey 08544, USA}
\date{\today}
\begin{abstract}
Alkali-metal atomic magnetometers suffer from heading errors in geomagnetic fields as the measured magnetic field depends on the orientation of the sensor with respect to the field. In addition to the nonlinear Zeeman splitting, the difference between Zeeman resonances in the two hyperfine ground states can also generate heading errors depending on initial spin polarization. We examine heading errors in an all-optical scalar magnetometer that uses free precession of polarized $^{87}\text{Rb}$ atoms by varying the direction and magnitude of the magnetic field at different spin polarization regimes. In the high polarization limit where the lower hyperfine ground state $F = 1$ is almost depopulated, we show that heading errors can be corrected with an analytical expression, reducing the errors by two orders of magnitude in Earth's field. We also verify the linearity of the measured Zeeman precession frequency with the magnetic field. With lower spin polarization, we find that the splitting of the Zeeman resonances for the two hyperfine states causes beating in the precession signals and nonlinearity of the measured precession frequency with the magnetic field. We correct for the frequency shifts by using the unique probe geometry where two orthogonal probe beams measure opposite relative phases between the two hyperfine states during the spin precession. 
\end{abstract}

\maketitle
%
\section{Introduction}

Total-field atomic magnetometers measure the magnitude of the magnetic field by directly measuring the Larmor precession frequency of the electron spins of alkali-metal atoms in the presence of the field. They can operate in geomagnetic fields (10 - 100 $\mu$T) and have a wide range of applications, including space magnetometry \cite{Korth2016,Sabaka2016,Friis-Christensen2006,Dougherty2006,Dougherty2004,Reigber2002}, fundamental physics experiments \cite{Abel2020,Lee2018,Altarev2009,Vasilakis2009}, biomedical imaging \cite{Limes2020,Zhang2020,Bison_2009,Bison2003}, archaeological mapping \cite{Linford2019,Linford2007}, mineral exploration \cite{Gavazzi2020,Walter2019,Proutya,Nabighian2005}, searches for unexploded ordnance \cite{Paoletti2019,Billings2006,Zhang2003,Nelson2001}, and magnetic navigation \cite{Canciani2017,Fu2020,Bevan2018,Shockley2014,Goldenberga}. The highest sensitivity for scalar magnetometers has been achieved in a pulsed pump-probe arrangement with a  sensitivity of 0.54 fT$/\sqrt{\text{Hz}}$ in a field of 7.3 $\mu$T \cite{Sheng2013}. However, practical magnetometers need to operate in geomagnetic field around 50 $\mu$T. Recently an all-optical pulsed gradiometer has reached a magnetometer sensitivity of 14 $\text{fT}/ \sqrt{\text{Hz}}$ over a broad range including Earth's field \cite{Lucivero2019}. One major and practical challenge of Earth's field magnetometers is the control of heading errors which otherwise significantly limit their accuracy. They cause the measured field values to depend on the orientation of the sensor with respect to the field, especially presenting problems for the magnetometry-based navigation \cite{Oelsner2019,Ben-Kish2010,Hovdea,Alexandrov2003}.

All alkali-metal magnetometers suffer from heading errors because alkali-metal isotopes have nonzero nuclear spin of $I>1/2$. There are mainly two physical sources of heading errors: the nonlinear Zeeman splitting due to mixing of ground Zeeman states $\left| F,m \right>$ and the difference in Larmor frequencies for the two hyperfine manifolds due to the nuclear magnetic moment. The non-linear splitting corresponds to a difference of 2.6 nT between neighboring Zeeman states for $^{87}$Rb in a 50 $\mu$T field. At this field, the linear difference between Zeeman resonance frequencies in $F=1$ and $F=2$ states is 200 nT. These splittings of the Zeeman resonance lines produce broadening and asymmetries in the lineshape depending on the orientation of the sensor with respect to the field. For $^{87}\text{Rb}$ in a 50 $\mu$T field, the orientation-dependent shifts are on the order of 15 nT. Previous approaches of reducing the heading errors in other alkali vapor systems have focused on suppressing the nonlinear Zeeman splitting, including double-modulated synchronous optical pumping \cite{Seltzer2007}, light polarization modulation \cite{Oelsner2019,Ben-Kish2010}, measurements of high-order polarization moments \cite{Acosta2008,Acosta2006a,Pustelny2006,Yashchuk2003}, use of tensor light-shift to cancel quadratic Zeeman splitting \cite{Jensen2009}, and spin-locking with an additional radio-frequency (RF) field \cite{Bao2018}. However, most of them have some practical drawbacks such as complexity in implementation or requiring use of RF fields. These methods also do not cancel frequency shifts associated with the difference of Zeeman resonances for $F=1$ and $F=2$ states. In magnetometers operated with continuous optical pumping, the optimal sensitivity is achieved for spin polarization generally near $50 \%$. As a result, there is usually a significant population in $F=1$ state which changes depending on the orientation of the magnetometer relative to the magnetic field.

In this paper, we study heading errors as a function of both the direction and magnitude of magnetic field at a wide range of initial spin polarization and implement several methods of their correction. With the all-optical, free-precession $^{87}\text{Rb}$ magnetometer, we use short-pulse pumping technique to achieve very high initial spin polarization near $95 \%$ regardless of the field orientation such that the initial spin state is well-defined. The population of $F=1$ state becomes negligible, and Zeeman coherences decay much faster in $F=1$ state than in $F=2$ state due to spin-exchange collisions between alkali-metal atoms. In this high polarization limit, we can minimize the polarization-dependent heading errors. We also find that the average Larmor frequency is given by a simple analytical expression that depends on the angle between the pump laser and the magnetic field. We show this angle can be determined directly from the spin precession signals. Thus one can calculate a correction for the heading error in real time. We verify the accuracy of the magnetometer as a function of both the magnitude and direction of the magnetic field with an accuracy of about 0.1 nT, suppressing heading errors by two orders of magnitude in a 50 $\mu$T field. 

At lower spin polarization we observe interesting effects due to non-negligible contribution from the $F = 1$ state. The difference in Zeeman frequencies of $F = 1$ and $F = 2$ states generate beating which is observable in the measurement of spin precession signals. Moreover, the measured spin precession frequency is no longer linear with the magnetic field, even though the splitting itself is linear with the field. Here we use two probe beams to further correct for these heading errors: one is collinear to the pump beam, and the other is perpendicular to the pump. These orthogonal probe beams measure opposite relative phases of the two hyperfine ground states during their precession, allowing one to cancel any effects from the splitting in their Larmor frequency by averaging the two probe measurements. This is due to the fact that Zeeman coherences precess around the magnetic field in opposite directions for $F=1$ and $F=2$ states. As a result, we cancel the additional frequency shifts by averaging the measurements of the two orthogonal probes. Furthermore, we compare our experimental results with a density matrix simulation to easily separate signals from $F=1$ and $F=2$ states and investigate frequency shifts due to the nuclear magnetic moment. 

\section{Analytical correction of heading errors}\label{sec:Theory}

For $^{87} \text{Rb}$ atoms in ground states with electronic spin $S = 1/2$, the energy of the Zeeman sublevel $\left| m \right>$ with total atomic angular momentum $F$ is given by the Breit-Rabi formula \cite{Breit1931}: 
\begin{equation}\label{eq:BreitRabi}
E = - \frac{\hbar \omega_\text{hf}}{2(2I+1)} - g_I \mu_B B m \pm \frac{\hbar \omega_\text{hf}}{2} \sqrt{x^2 + \frac{4mx}{2I+1}+1}.
\end{equation}
where $x = (g_s+ g_I)\mu_B B/\hbar \omega_{\text{hf}}$, $g_s$ and $g_I = \mu_I/(\mu_B I)$ are the electronic and nuclear Land$\acute{\text{e}}$ factors respectively, $\mu_B$ is the Bohr magneton, $B$ is the magnetic field strength, $\omega_\text{hf}$ is the hyperfine splitting, $I$ is the nuclear spin, and the $\pm$ refers to the $F = I \pm 1/2$ hyperfine components. In the Earth-field range, the $m \rightarrow m-1$ Zeeman transition frequency is given by \cite{Seltzer2007}: 
\begin{align}\label{eq:transitionfA}
\omega_{F, m} & \simeq \frac{(\pm \mu_{\text{eff}}-g_I \mu_B)B}{\hbar} \mp \frac{\mu^2_{\text{eff}}B^2}{\hbar^2 \omega_{hf}}(2m - 1) \nonumber \\
			& = \omega_L \mp \omega_\text{rev} (2m -1)
\end{align}
where $\mu_\text{eff} = (g_s \mu_B + g_I \mu_B)/ (2I+1)$, $\omega_L = (\pm \mu_\text{eff}-g_I\mu_B)B/\hbar$ is the Larmor frequency, and $\omega_\text{rev} = \mu^2_{\text{eff}}B^2/\hbar^2 \omega_{hf} $ is the quantum-beat revival frequency which is nonlinear to the field magnitude. The Larmor frequencies for the two hyperfine states are approximately opposite, but not exactly equal because of the $g_I \mu_B$ term. The difference of absolute frequencies is proportional to the magnetic field and equal to 1.4 kHz at 50 $\mu$T, where the Larmor frequency is equal to 350 kHz for $^{87}$Rb atoms. The nonlinear Zeeman effect in Earth's field causes a splitting of 18 Hz between neighboring Zeeman transitions, which is non-negligible for magnetometer operation. 

The measured transverse spin component can be written in terms of the $^{87}$Rb ground state density matrix as a weighted sum of coherences oscillating at different Zeeman frequencies, given by
\begin{equation}\label{eq:TransSpin}
\left< S_x \right> = \text{Tr}\left[\rho S_x \right] = \sum_{F \,m' = m \pm 1} A_{F\,m,m'} \rho_{F\,m,m'}
\end{equation}
where $\rho_{F\,m,m'}$ is the off-diagonal element of density matrix for an ensemble of $^{87} \text{Rb}$ atoms in coupled basis $\left|F\,m \right>$, and $A_{F\,m,m'}$ is its amplitude. We leave a detailed discussion of the density matrix analysis to Sec.~\ref{sec:Appendix-2} of the Appendix. The measured spin precession frequency is therefore a combination of different Zeeman transition frequencies. Any variation in sensor's orientation with respect to the field can change the relative strength between the coherences, shifting the measured precession frequency. 

In the high spin polarization limit we derive in Sec.~\ref{sec:Appendix-3} of the Appendix the heading error correction:
\begin{equation}\label{eq:headingerrorcorrectionA}
B = \frac{4h\nu}{(g_S - 3 g_I)\mu_B}\left[1 - \frac{3\nu}{\nu_\text{hf}}\text{sin}\theta \frac{P(7+P^2)}{5+3P^2}\right]
\end{equation}
where $\nu = \omega/2\pi$ is the measured precession frequency, $\nu_\text{hf}$ is the hyperfine splitting frequency, $P$ is degree of initial spin polarization, and $\theta$ is the angular deviation of the pump beam from the nominal magnetometer orientation where the pump laser is perpendicular to the magnetic field (see Sec.~\ref{sec:magnetometerdes} for more details). We assume the relative distribution of atoms in $F=2$ state is given by spin-temperature distribution. The spin temperature distribution is realized when the rate of spin-exchange collisions is higher than other relaxation rates \cite{Anderson1960}. It is also realized during optical pumping on a pressure broadened optical resonance with fast $J$-damping in the excited state \cite{Appelt1998}. These conditions are reasonably well-satisfied in our experiment. In a $50\,\mu \text{T}$ Earth's field, the maximum size of the correction given by Eq.~\ref{eq:headingerrorcorrectionA} is on the order of 15 nT with full polarization ($P = 1$). 

\begin{figure}
\centering
\includegraphics[width=0.9\columnwidth]{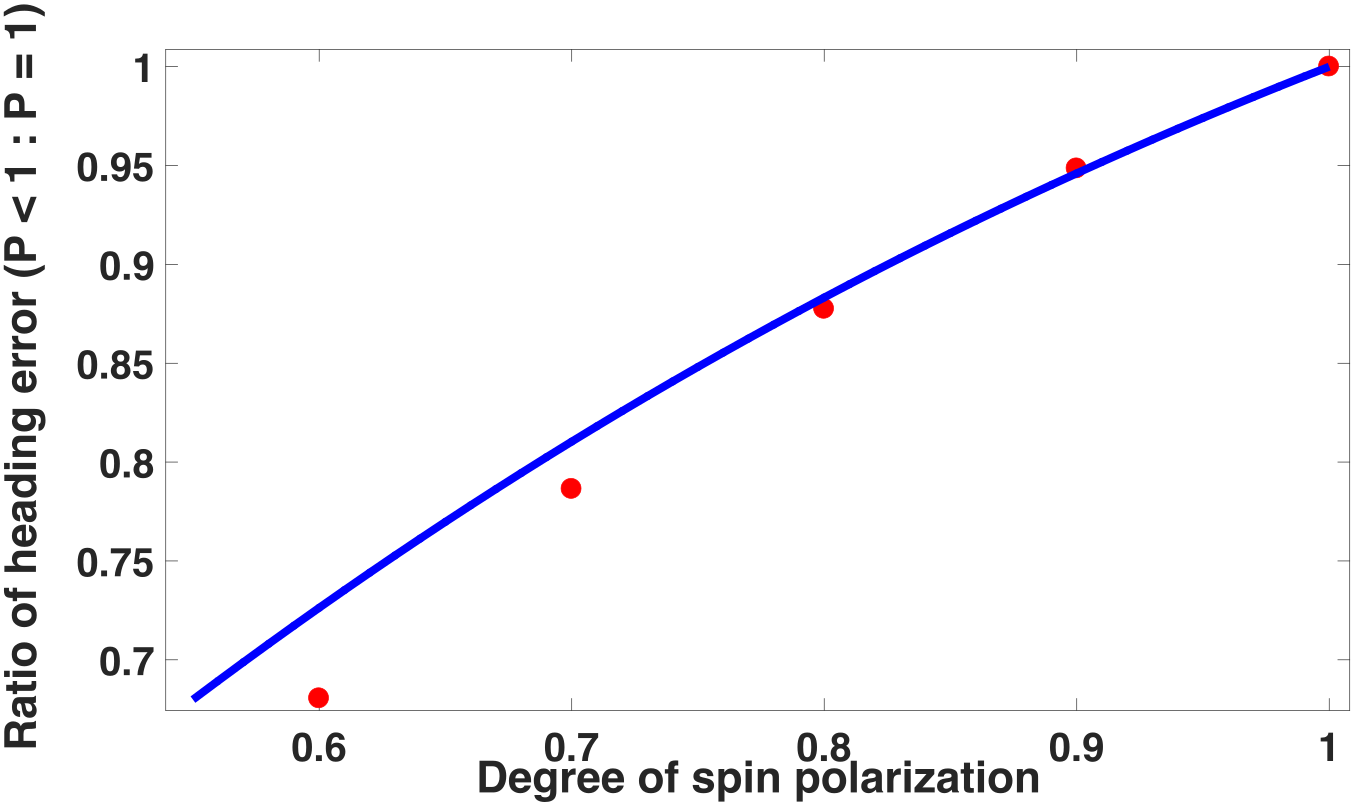}
\caption{The ratio of heading error correction at spin polarization $P<1$ to $P = 1$. The theoretical prediction calculated from Eq.~\ref{eq:headingerrorcorrectionA} (blue line) and the numerical simulation result (red circles) are compared.}
\label{fig:ComparisonAnalyticalSimulation.png}
\end{figure}

Fig.~\ref{fig:ComparisonAnalyticalSimulation.png} shows the comparison between heading error correction calculated with Eq.~\ref{eq:headingerrorcorrectionA} and numerical simulation of the density matrix evolution. The density matrix model, described in more detail in Sec.~\ref{sec:Appendix-4} of the Appendix, includes optical pumping, free spin precession evolution, and signal fitting as done in the experiment. They agree well in high polarization limit. Thus, if we experimentally find the initial polarization and extract the angle $\theta$, we can use Eq.~\ref{eq:headingerrorcorrectionA} to find the heading-error free magnetic field. At low polarization more complex behavior is expected due to the incomplete depopulation of $F = 1$ state. 

\section{A pulsed-pump double-probe $^{87}\text{RB}$ magnetometer}\label{sec:magnetometerdes}

\begin{figure}
\centering
\includegraphics[width=0.9\columnwidth]{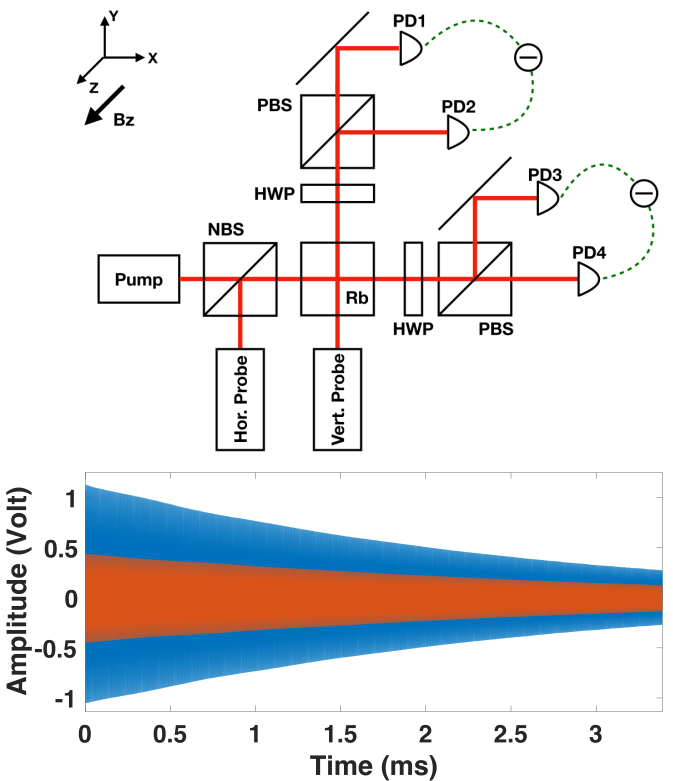}
\caption{Top: Schematic of an integrated $^{87}\text{Rb}$ magnetometer. NBS, nonpolarizing beam splitter; HWP, half-waveplate; PBS, polarizing beam splitter; PD, photodiode. Bottom: Experimental rotation signals of horizontal probe (blue) and vertical probe (red) acquired from the digital oscilloscope.}
\label{fig:Schematic_SignalsA90.png}
\end{figure}

We use a compact integrated magnetometer with schematic shown in Fig.~\ref{fig:Schematic_SignalsA90.png}. It consists of a $^{87}\text{Rb}$ vapor cell, electric heaters, a pump laser, two probe lasers, and two polarimeters. The cell has height and width of 5 mm, and length of 10 mm. It contains internal mirrors that allows pump and probe beams to reflect back and forth many times inside the cell. The probe beams exit the cell after 11 passes. The sensor is placed inside a magnetic shielding on a rotational stage that allows rotation of the whole assembly relative to the magnetic field. The multipass cell is filled with enriched $^{87}\text{Rb}$ vapor and 700 Torr $\text{N}_2$ buffer gas. It is typically heated to 100$^\circ$C giving a $^{87}\text{Rb}$ number density of $4\times10^{12}\,/\text{cm}^3$ as estimated based on measured transverse spin relaxation rate which is further described in Sec.~\ref{sec:Appendix-2} of the Appendix. 

The pump laser is directed in $x$ direction in Fig.~\ref{fig:Schematic_SignalsA90.png}, circularly polarized and tuned to the $^{87}\text{Rb}$ D1 line. It is operated in a pulsed mode with several watts instantaneous power per microsecond pulse. It has an adjustable repetition rate, number, and width of pulses. The usual pumping cycle consists of 190 pulses with a 70 ns single pulse width that is much shorter than the Larmor period $\tau = 2\pi/\gamma B_z$. For a resonant build-up of the spin polarization, the pulse repetition rate is synchronous with the Larmor frequency. The magnetic field is created by a concentric set of cylindrical coils inside two layers of $\mu$-metal magnetic shields. The spin free precession is detected by two off-resonant vertical-cavity surface-emitting laser (VCSEL) probe beams which are linearly polarized. One is in $x$ direction, along the pump propagation (``horizontal probe''), and the other is in $y$ direction, orthogonal to the pump (``vertical probe''). Each balanced polarimeter consists of a half-wave plate, a polarizing beamsplitter, and two photodiodes with differential amplification. The two differential signals $V_\text{ver} = V_1 - V_2$ and $V_\text{hor} = V_3 - V_4$ are recorded by a digital oscilloscope and have the general form of a sine wave with exponential decay plus an offset: 
\begin{equation}\label{eq:ExpDecaySine}
V(t) = V_0\, \text{cos} (2\pi\nu t + d) e^{-t/T_2} + V_\text{DC}(t)
\end{equation}
where $V_0$ is the initial amplitude, $\nu$ is the precession frequency, $d$ is the phase delay, $T_2$ is the transverse spin relaxation time, and $V_\text{DC}$ is the offset. The lower panel of Fig.~\ref{fig:Schematic_SignalsA90.png} shows the experimentally measured signals at 50 $\mu$T. The vertical probe shows a smaller signal than the horizontal probe as it has a smaller interaction volume from a smaller overlap region with the pump beam. An external frequency counter (HP53310A) measures the frequency of the signal by detecting its zero-crossings during the free precession measurement time $T_m = 3$ ms, which is comparable to $T_2$. We add external high-pass filters with $f_c = 20\, \text{kHz}$ to cancel DC offsets in the signals going to the frequency counter. We continuously measure the frequency and read the center frequency of its histogram distribution with a standard deviation of about 1 mHz. 

\begin{figure}
\centering
\includegraphics[width=0.7\columnwidth]{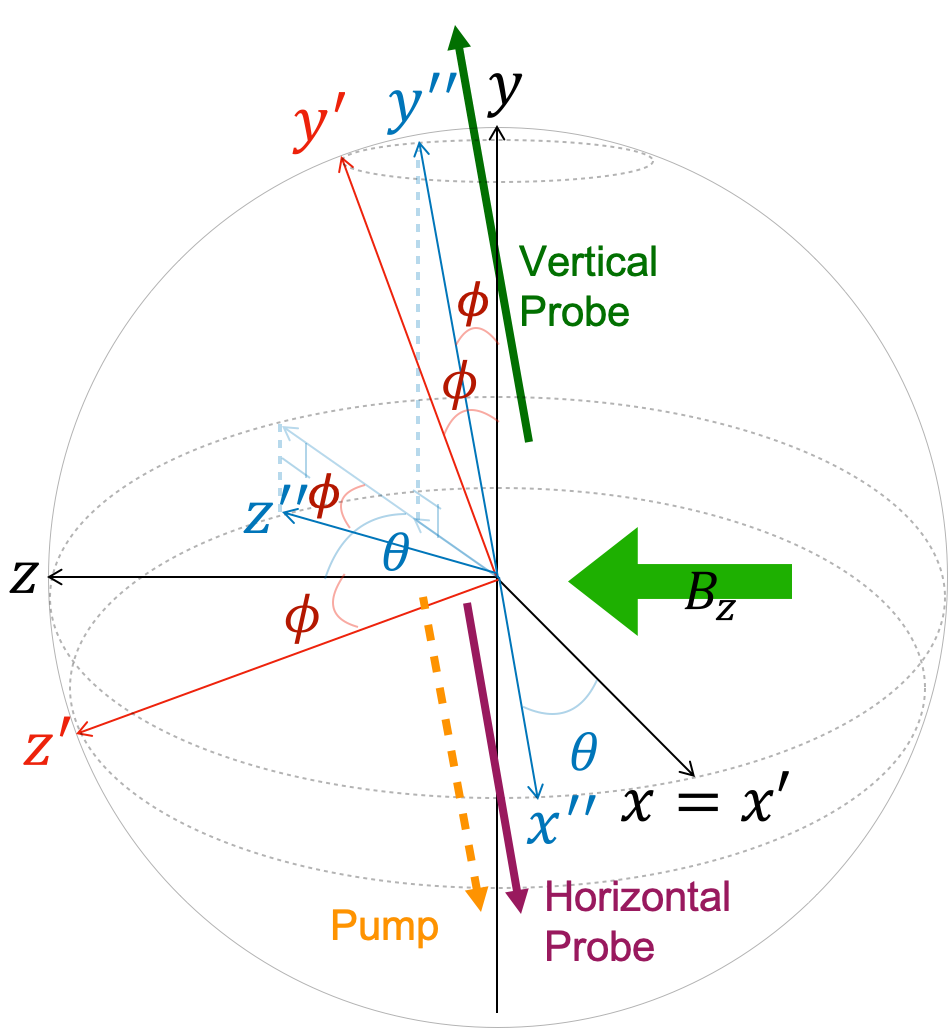}
\caption{Pump-probe geometry of a $^{87} \text{Rb}$ magnetometer. Initially the pump and the horizontal probe beams are in $x$; the vertical probe beam is in $y$; the magnetic field is in $z$. If the sensor gets tilted, we apply two successive rotations: rotation by $\phi$ about $x$ to $x'y'z'$ system and rotation by $\theta$ about $y$ to $x''y''z''$ system. In the final configuration, the pump and the horizontal probe beams are in $x''$, and the vertical probe beam is in $y''$.}
\label{fig:MagnetometerBasics.pdf}
\end{figure}

Fig.~\ref{fig:MagnetometerBasics.pdf} describes the pump-probe geometry after a change in sensor orientation. The horizontal probe is always collinear to the pump, and the vertical probe is always orthogonal to the pump. In the initial configuration ($\theta = 0^\circ$), the pump and horizontal probe are in $x$, and the vertical probe is in $y$. After a sensor rotation, the pump beam and horizontal probe are in $x''$ direction and tilted by $\theta$ from the initial magnetometer orientation $x$ where the field $B_z$ is perpendicular to the spin. The vertical probe is in $y''$ direction at a small angle $\phi = 11^\circ$ from $y$. With the tilt of the sensor, the rotation matrix transforming $x,y,z$ to $x'',y'',z''$ coordinates is
\begin{align}\label{eq:pump-probe geometry}
R & = 
\begin{pmatrix}
\text{cos}\theta & 0 & -\text{sin}\theta \\
0 & 1 & 0 \\
\text{sin}\theta & 0 & \text{cos}\theta 
\end{pmatrix}
\begin{pmatrix}
1 & 0 & 0 \\
0 & \text{cos}\phi & -\text{sin}\phi \\
0 & \text{sin}\phi & \text{cos}\phi 
\end{pmatrix} \nonumber \\
& = 
\begin{pmatrix}
\text{cos}\theta & -\text{sin}\theta\,\text{sin}\phi & -\text{sin}\theta\,\text{cos}\phi \\
0 & \text{cos}\phi & -\text{sin}\phi \\
\text{sin}\theta & \text{cos}\theta\,\text{sin}\phi & \text{cos}\theta\,\text{cos}\phi
\end{pmatrix}
.
\end{align}
In the final configuration, the pump and the horizontal probe are therefore in $\hat{x}'' = \text{cos}\theta\,\hat{x} + \text{sin}\theta\,\hat{z}$, and the vertical probe is in $\hat{y}'' = -\text{sin}\theta\,\text{sin}\phi\,\hat{x} + \text{cos}\phi\,\hat{y} + \text{cos}\theta\,\text{sin}\phi \,\hat{z}$.

\section{Measurement of heading errors}\label{sec:procedure}

\begin{figure}
\centering
\includegraphics[width=0.9\columnwidth]{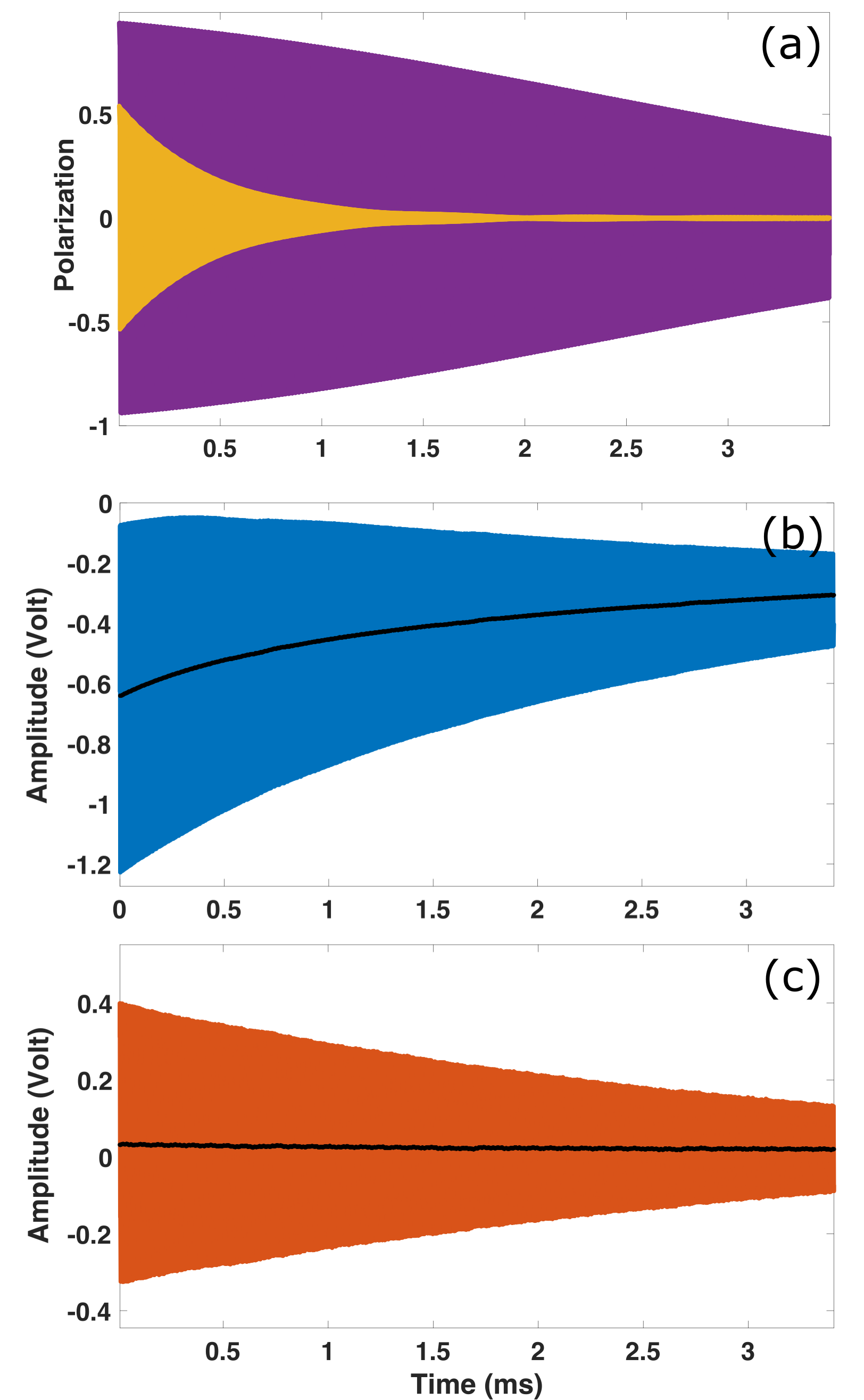}
\caption{(a) Simulated decay for $P = 2 \left< S_x \right>$ with initial polarization $P_0 = 0.95$ for $F = 1$ signal $\times 100$ (yellow) and $F = 2$ signal (violet). (b) Measured spin precession signal in $50 \,\mu$\textnormal{T} at $\theta = 45^{\circ}$ by the horizontal probe. (c) Measured signal under the same condition by the vertical probe.}
\label{fig:F1VsF2_P095A90.png}
\end{figure}

The pulsed pump laser can achieve very high initial spin polarization near $95 \%$. This minimizes the polarization-dependent heading errors. In Fig.~\ref{fig:F1VsF2_P095A90.png}(a) we plot the simulated signals for each hyperfine state separately, showing how much each hyperfine state contributes to $\left< S_x \right>$ for initial atomic spin polarization $P = 0.95$. The $F = 1$ signal has a very small initial amplitude compared to the $F = 2$ signal and decays faster during the precession. To maximize the polarization experimentally, we adjust the width and number of pump pulses until the initial signal amplitude saturates.

In order to apply Eq.~\ref{eq:headingerrorcorrectionA} to correct for the heading error one must know the tilt angle $\theta$ of the magnetometer as shown in Fig.~\ref{fig:MagnetometerBasics.pdf}. In the absence of other information about the field direction one can find the angle $\theta$ from the signal itself by considering the DC component of the spin precession signal. Fig.~\ref{fig:F1VsF2_P095A90.png}(b) and \ref{fig:F1VsF2_P095A90.png}(c) show the measured signals at $\theta = 45^\circ$ which have nonzero time-varying DC offsets compared to Fig.~\ref{fig:Schematic_SignalsA90.png}. The DC offset measures the spin component parallel to the magnetic field. As shown in Fig.~\ref{fig:MagnetometerBasics.pdf}, the sensor has a small tilt about $x$ by $\phi$ such that the vertical probe is not perfectly transverse to the field $B_z$. As a result, the vertical probe signal also gains a small DC offset. 

\begin{figure}
\centering
\includegraphics[width=\columnwidth]{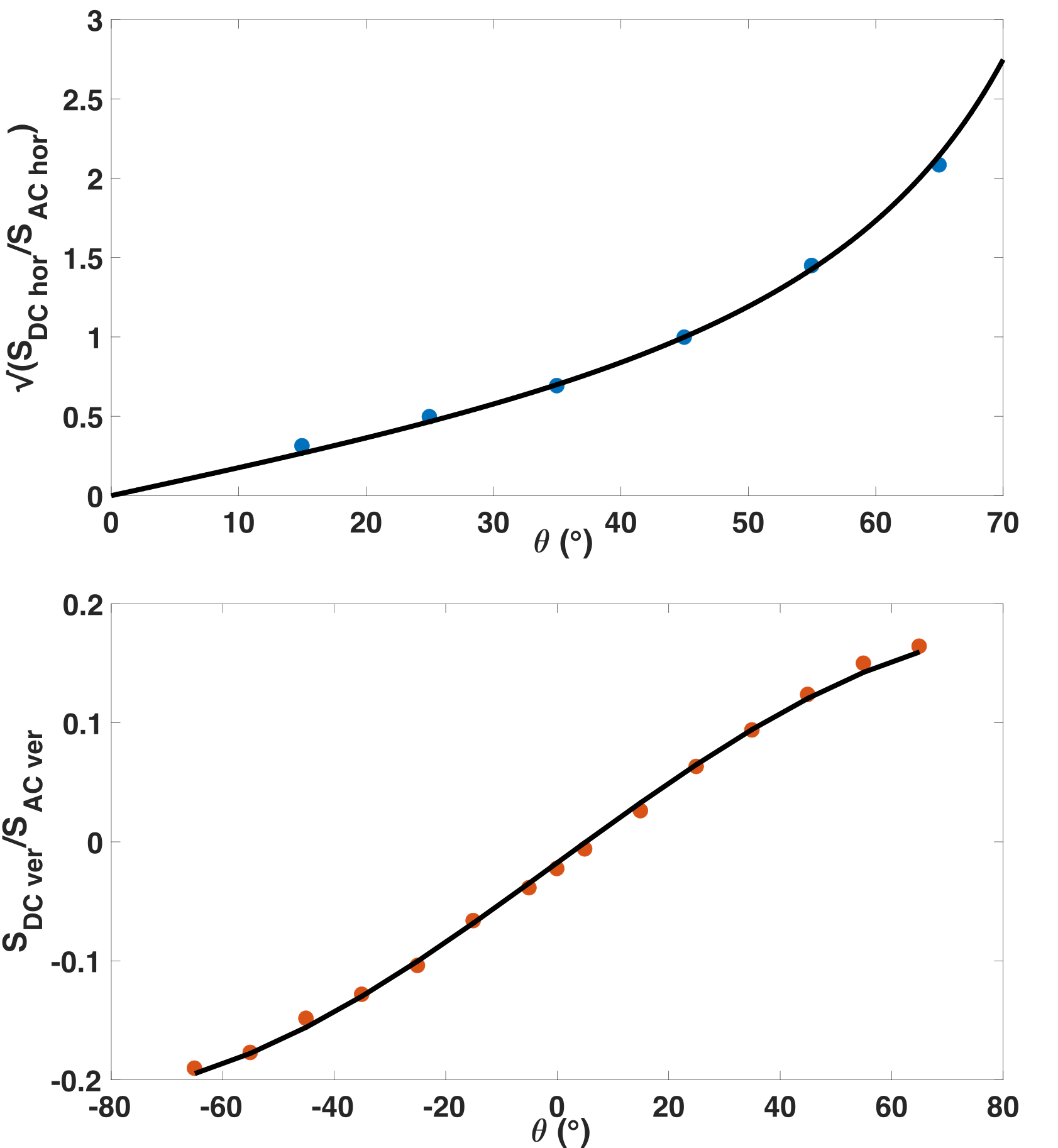}
\caption{Measured initial ratio of DC offset to AC amplitude for the horizontal probe (top) and vertical probe (bottom) signals as a function of $\theta$. For the horizontal probe, it is fit to: $\sqrt{\text{DC}/\text{AC}} = \text{tan}\,\theta$. For the vertical probe, it is fit to: $\text{DC}/\text{AC} =\sin\theta \sin\phi/ \sqrt{1-\cos^2\theta \sin^2 \phi} + a$, where $\phi = 11.2^\circ$ is the tilt angle of the vertical probe (see Fig.~\ref{fig:MagnetometerBasics.pdf}) and $a = -0.018$ is the offset when $\theta = 0^\circ$. }
\label{fig:ACDC.png}
\end{figure}

From Eq.~\ref{eq:pump-probe geometry}, the initial optically pumped spin is $\vec{S} = S\hat{x}'' = S\text{cos}\theta\,\hat{x} + S\text{sin}\theta\,\hat{z}$. Ignoring the spin relaxation for simplicity, the precessing spin at angular velocity $\omega$ is then $\vec{S} = S\text{cos}\theta\,\text{cos}\omega t\,\hat{x} + S\text{cos}\theta\,\text{sin}\omega t\,\hat{y}  + S\text{sin}\theta\,\hat{z}$. The horizontal probe detects the spin component 
\begin{equation}
S_{x''} = \vec{S}\cdot\hat{x}'' = S_\text{AC}+S_\text{DC} = S\text{cos}^2\theta\, \text{cos}\omega t+ S\text{sin}^2\theta. 
\end{equation}
The first term is the projection of $S_x$ component which oscillates. The second term is the projection of $S_z$ component, resulting in the DC offset. Therefore, the ratio of the initial DC offset to maximum AC amplitude is $S_\text{DC}/S_\text{AC} = \text{tan}^2\theta$. We show in Fig.~\ref{fig:ACDC.png} good agreement between this equation and our measurements, allowing us to estimate the magnitude of $\theta$.

We can also determine the magnitude of $\phi$ based on the vertical probe signal. The vertical probe detects the spin component 
\begin{align}
& S_{y''} = \vec{S}\cdot\hat{y}'' = S_\text{AC} + S_\text{DC} \\ 
& = S\text{cos}\theta \left(\text{cos}\phi \, \text{sin}\omega t - \text{sin}\theta\, \text{sin}\phi\, \text{cos}\omega t \right) + S\text{sin}\theta\,\text{cos}\theta\,\text{sin}\phi. \nonumber
\end{align}
The ratio of the initial DC offset to maximum AC amplitude is then $S_{\text{DC}}/S_{\text{AC}}= \sin\theta \sin\phi/ \sqrt{1-\cos^2\theta \sin^2 \phi}$. The lower panel of Fig.~\ref{fig:ACDC.png} shows the measurement of the $S_\text{DC}/S_\text{AC}$ ratio of the vertical probe signal, which gives an estimation of $\phi = 11.2^\circ$. We cannot find the sign of $\theta$ and $\phi$ independently since they are coupled as shown in the expression of $S_{\text{DC}}/S_{\text{AC}}$. 

To measure the heading errors we tilt the sensor with respect to the field in the range $ -65^{\circ} < \theta  < +65^{\circ}$ and measure the spin precession frequency with the frequency counter. It is important to separate heading errors due to spin interactions from heading errors associated with remnant magnetization of magnetometer components. Rotation of the sensor relative to the field changes the projection of the remnant magnetic fields onto the leading field, resulting in frequency shifts that are hard to distinguish from atomic heading errors. The sensor was constructed with a minimal number of magnetic components. However, there are small amounts of polarizable ferrous materials present in the laser mounts and other electronic components. We have degaussed these components and turned off heater electric currents during the measurement. Nevertheless, small offsets on the order of a few nT due to remnant magnetization of the sensor remained. To account for these offsets, we periodically reversed the polarization of the pump laser with a half-wave plate and took measurements with both polarizations. This method is often used to cancel heading errors by averaging the signals from the two pump polarizations \cite{Yabuzaki1974,Ben-Kish2010,Oelsner2019}. In this case we took the difference of the signals to separate the heading errors due to the spin interaction and those due to magnetization of the components in the sensor head.

\section{Heading errors as a function of the sensor orientation}

\begin{figure}
\centering
\includegraphics[width=\columnwidth]{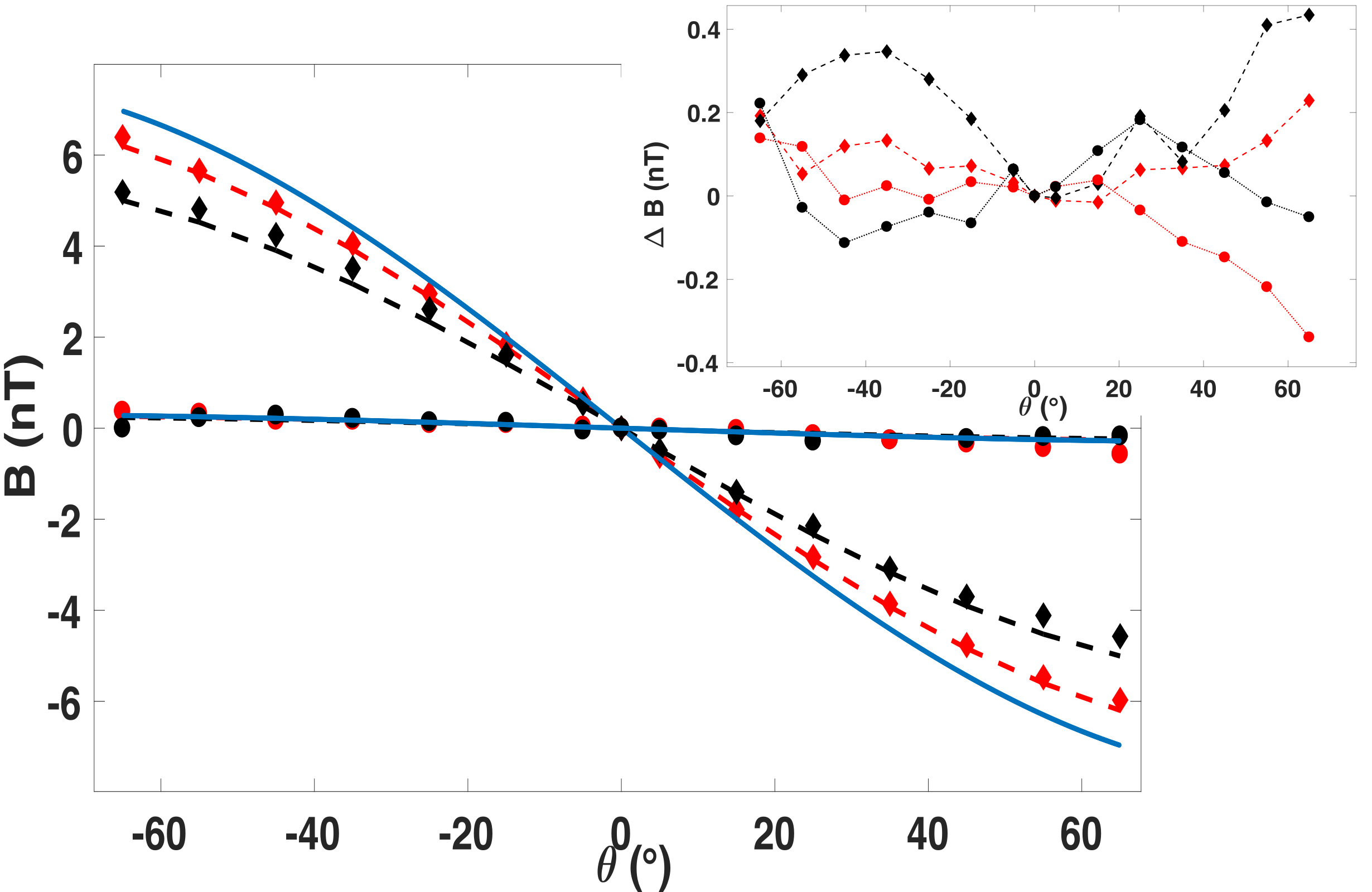}
\caption{Measurement of heading errors in $B = 10 \,\mu$\textnormal{T} (circles) and $50 \,\mu$\textnormal{T} (diamonds) as a function of $\theta$. They are measured from both horizontal (red) and vertical probe signals (black). Based on fitting to Eq.~\ref{eq:headingerrorcorrectionA} (dashed lines), the initial polarizations of the horizontal and vertical probe signals are estimated to be $P = 0.90$ and $P = 0.68$ respectively. The heading error estimation at full polarization $P = 1$ is also shown (blue continuous lines). Inset: The residuals of fitting of the heading errors. In $B = 10\, \mu\text{T}$ the vertical (black circles) and horizontal (red circles) probe measurements have residuals with standard deviation of 0.15 nT and 0.098 nT respectively. In $B = 50\, \mu\text{T}$ the vertical (black diamonds) and horizontal (red diamonds) probe measurements have residuals with standard deviation of 0.12 nT and 0.07 nT respectively. }
\label{fig:B05G01GBvsA_Residuals.png}
\end{figure}

Fig.~\ref{fig:B05G01GBvsA_Residuals.png} reports the measurement of heading errors in $10\, \mu\text{T}$ and $50 \,\mu\text{T}$ fields as a function of the sensor's tilt angle $\theta$. If we take a difference between the two measurements at opposite pump polarizations, the field values at $\theta = 0^{\circ}$ are canceled to the order of 0.01 nT. The angle $\theta$ is determined from spin precession signals with an uncertainty of about 1$^\circ$. Fitting to Eq.~\ref{eq:headingerrorcorrectionA} gives an estimation of the initial polarization, about $P = 0.9$ for the parallel probe signal and $P = 0.7$ for the vertical probe signal. As described previously, the vertical probe has a smaller overlap with the pump beam than the horizontal probe and measures optical rotation from atoms that are less polarized. In a $10\, \mu\text{T}$ field, the heading errors are expected to be only $4\%$ of those in $50\, \mu\text{T}$. The inset of Fig.~\ref{fig:B05G01GBvsA_Residuals.png} shows that the residual errors are comparable and in fact slightly smaller for the case of B = 50 $\mu$T. This indicates that the residuals are likely due to imperfection in the cancellation of remnant magnetization of sensor components. Even so, we show that the heading errors due to atomic physics effects are reduced by about two orders of magnitude by the correction given by Eq.~\ref{eq:headingerrorcorrectionA}. 

\begin{figure}
\centering
\includegraphics[width=\columnwidth]{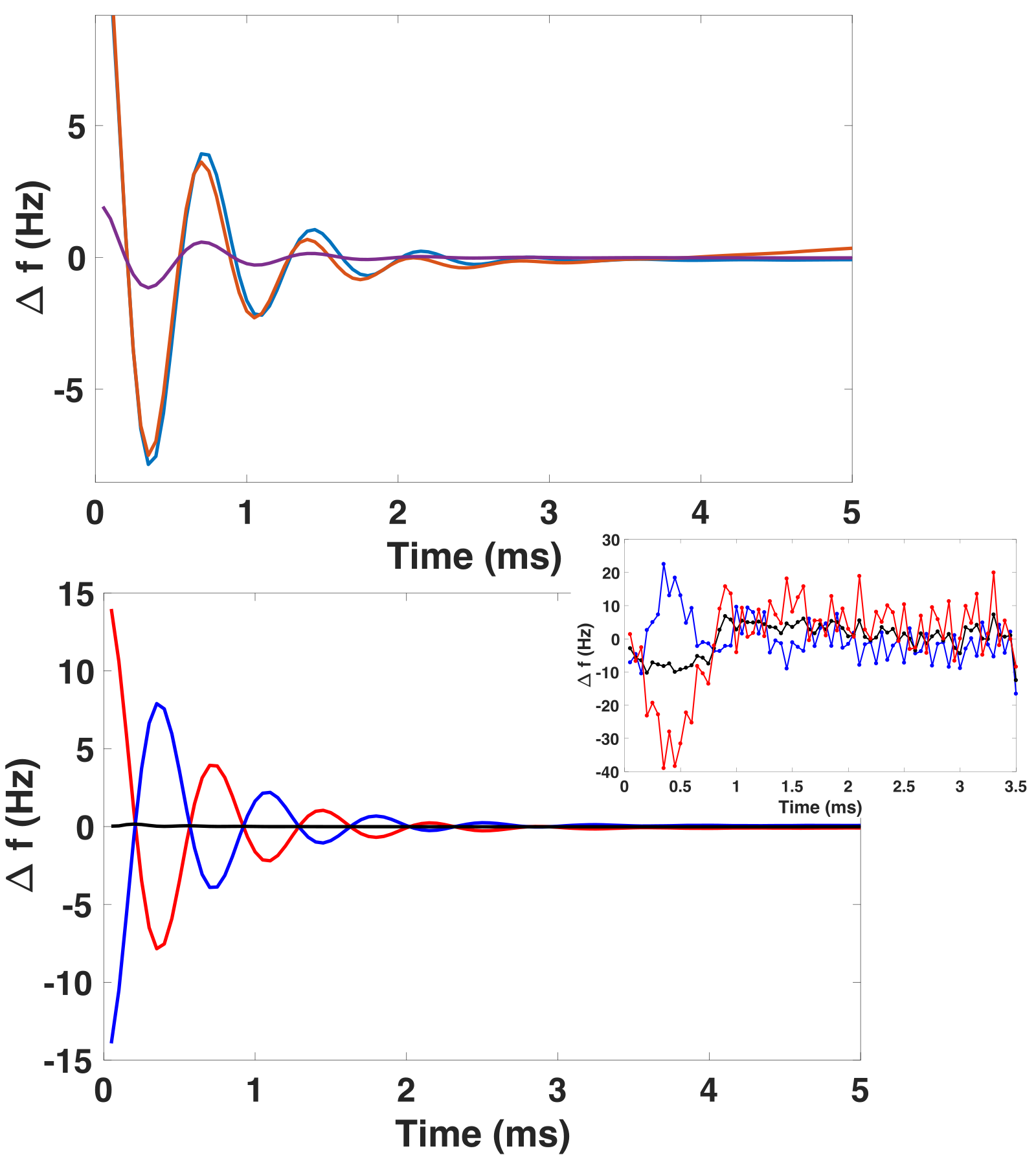}
\caption{Top: Simulated time-evolution of precession frequency in $B = 50 \,\mu$\textnormal{T}. With $P = 0.8$ at $\theta = 0^{\circ}$ (blue) and at $\theta = 45^{\circ}$ (red), the initial oscillation amplitude is about $\Delta f_0 = 14$ Hz. With $P = 0.99$ at $\theta = 0^{\circ}$ (purple), it is $\Delta f_0 = 2$ Hz. The frequency is estimated by fitting the simulated signal to Eq.~\ref{eq:ExpDecaySine} with each time segment for 0.05 ms. Bottom: Simulated time-evolution of precession frequency with $P = 0.8$ at $\theta = 0^{\circ}$. The vertical probe (red) and horizontal probe (blue) results both have $\Delta f_0 = 14$ Hz, and their average (black) has $\Delta f_0 = 0.03$ Hz. Inset: Experimental result with each segment fitting time of 0.05 ms. The vertical probe measurement (red) has $\Delta f_0 = 38$ Hz, the horizontal probe measurement (blue) has $\Delta f_0 = 23$ Hz, and their average (black) has $\Delta f_0 = 10$ Hz. }
\label{fig:FTimeEvolution.png}
\end{figure}

When the initial spin polarization is less than unity, there is some contribution from $F = 1$ state. This manifests itself as an oscillation in the instantaneous spin precession frequency as illustrated in Fig.~\ref{fig:FTimeEvolution.png}. The simulated spin precession signals are fit to Eq.~\ref{eq:ExpDecaySine} in individual time segments of 0.05 ms to show the time dependence of the spin precession frequency. We find that it oscillates at 1.4 kHz, which is equal to the difference between Zeeman frequencies for $F=1$ and $F=2$ states at 50 $\mu$T. We confirmed that at other magnetic fields the beating frequency is proportional to the magnetic field. As expected, the amplitude of the oscillations becomes larger for smaller spin polarization. The decay rate of the oscillations depends on the $T_2$ of the $F=1$ coherences. This beating effect is not sensitive to the orientation of the sensor, so the oscillations at $\theta=0$ and $\theta=45^{\circ}$ are similar. However, for $\theta=45^{\circ}$ one can observe a small additional slow drift of the  spin precession frequency. This drift is due to differences in the relaxation rate of $F=2$ coherences, as discussed in more detail in section~\ref{Sec.VI}.

We find that the oscillations in the instanteneous spin precession frequency have opposite sign for the horizontal and vertical probe beams, as illustrated in the bottom panel of Fig.~\ref{fig:FTimeEvolution.png}. As the two hyperfine states have opposite spin precession directions, the horizontal probe detects maximum signal when $\langle S_x\rangle_{F=2}$ and $\langle S_x\rangle_{F=1}$ are out of phase while the vertical probe detects maximum signal when they are in phase. We can therefore cancel the frequency oscillation by averaging the two probe measurements, which reduces the amplitude of frequency oscillations by more than two orders of magnitude. This was experimentally verified as shown in the inset of the lower Fig.~\ref{fig:FTimeEvolution.png}, which is based on fitting the experimental signal to Eq.~\ref{eq:ExpDecaySine}. The two probe measurements show opposite sign of frequency oscillation. The oscillations are larger for the vertical probe beam because it detects a lower average spin polarization. The amplitudes of the experimentally observed oscillations in the instantaneous spin precession frequency appear larger than those predicted by the simulation. This could be due to spatial non-uniformity of the polarization in the cell which is not taken into account in the simulation.

\section{Heading errors as a function of the absolute magnetic field}\label{Sec.VI}

\begin{figure}
\centering
\includegraphics[width=0.8\columnwidth]{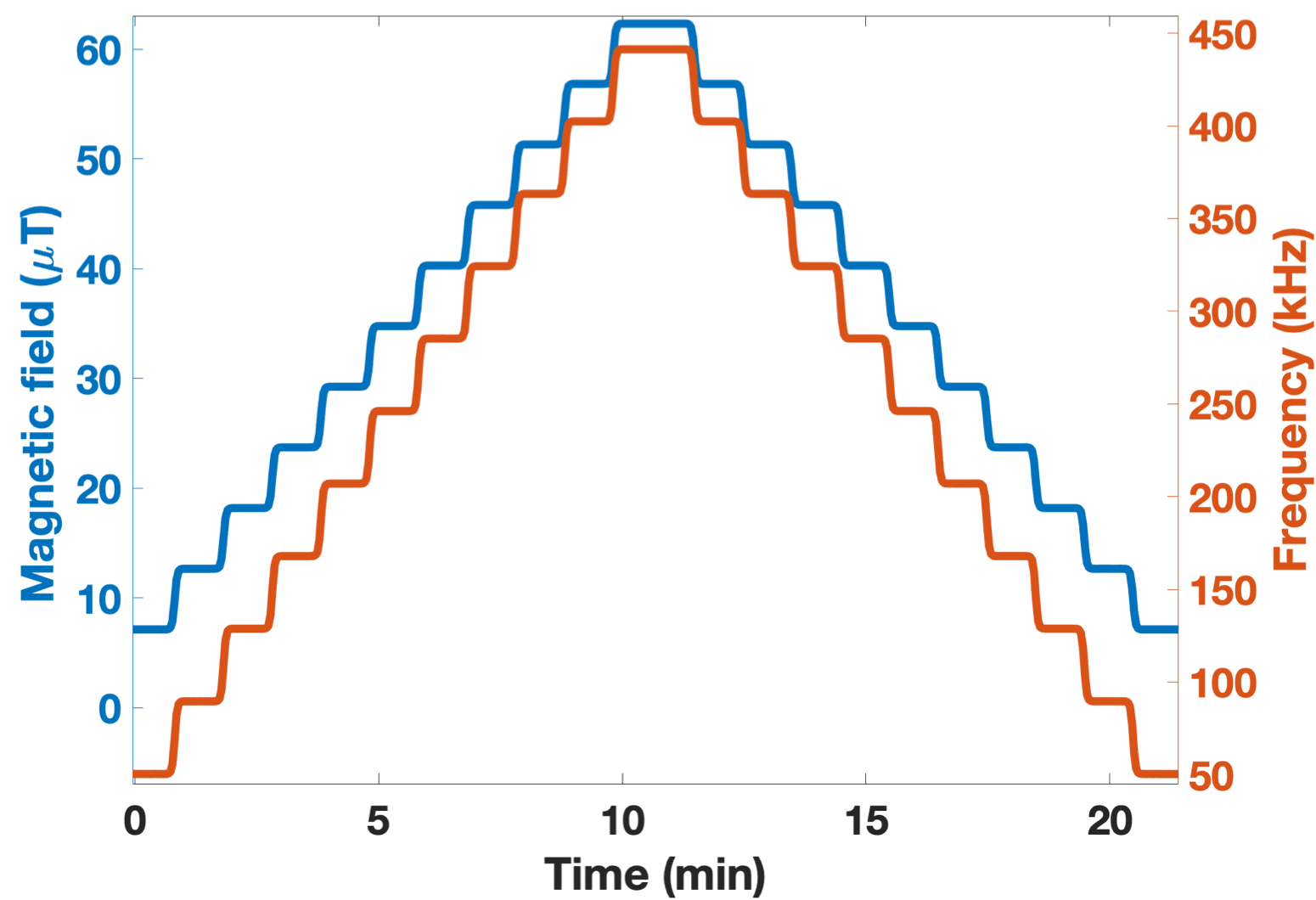}
\caption{Systematic modulation of magnetic field (blue) and simultaneous measurement of precession frequency (red). Each measurement lasts for a minute with transition time of 10 s.}
\label{fig:Bmodulation.png}
\end{figure}

In addition to investigating heading errors as a function of the angular orientation of the sensor, we also study them as a function of the absolute magnetic field. We measure the spin precession frequencies as a function of magnetic fields at high ($P = 0.85$) and low ($P = 0.2$) polarizations with the vertical probe. For these measurements it is necessary to create a well-controlled linear magnetic field ramp. The measurements are performed inside magnetic shields which generate a significant hysteresis of the magnetic field \cite{Feinberg2018}. To create a reproducible magnetic field we apply a stair-case ramp as illustrated in Fig.~\ref{fig:Bmodulation.png}. The magnetic field scan is repeated several times without interruptions. The current through the magnetic field coil is monitored using a precision shunt resistor while the repetition rate of the optical pumping pulses is continuously adjusted under computer control to match the Larmor frequency. After several up and down sweeps, we average both the field and frequency measurements during the same time period to suppress any field fluctuations. 

\begin{figure}
\centering
\includegraphics[width=\columnwidth]{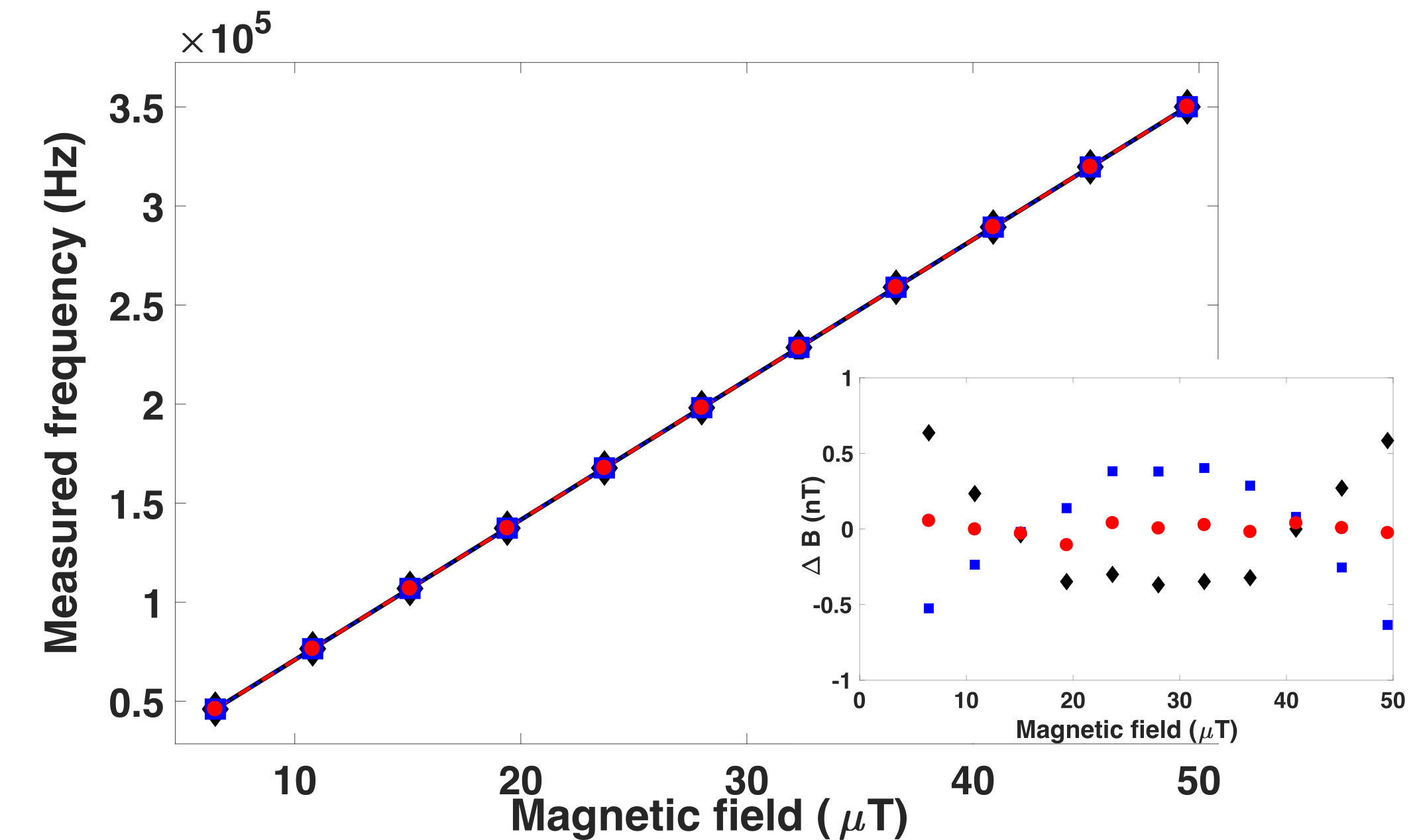}
\caption{The measurement of spin precession frequency as a function of fields with $P = 0.85$ at $\theta = 0^{\circ}$ with the vertical probe. It is fit to quadratic polynomial function (dashed lines). The curvature is $C = 153.99 \times 10^{-4} \text{Hz}/\mu \text{T}^2$ for the upward field sweep (black diamonds), $C = -149.09 \times 10^{-4} \text{Hz}/\mu \text{T}^2$ for the downward field sweep (blue squares), and $C = 2.46 \times 10^{-4} \text{Hz}/\mu \text{T}^2$ for the case of averaging the two measurements (red circles). Inset: The residuals of linear fitting of the frequency measurement as a function of fields, expressed in magnetic field unit. The standard deviation is 0.38 nT for the upward field sweep (black diamonds), 0.37 nT for the downward field sweep (blue squares), and 0.045 nT for the case of averaging the two measurements (red circles). }
\label{fig:FvsB_P085TotMeasurement.pdf}
\end{figure}

Fig.~\ref{fig:FvsB_P085TotMeasurement.pdf} shows the measured precession frequency as a function of fields with $P = 0.85$ at $\theta = 0^{\circ}$. With a quadratic fitting function, the frequency for upward and downward field sweeps show high curvature with opposite signs. This is due to the magnetic shield hysteresis, and averaging the two measurements reduces the curvature and residuals of linear fitting by two orders of magnitude. 

\begin{figure}
\centering
\includegraphics[width=\columnwidth]{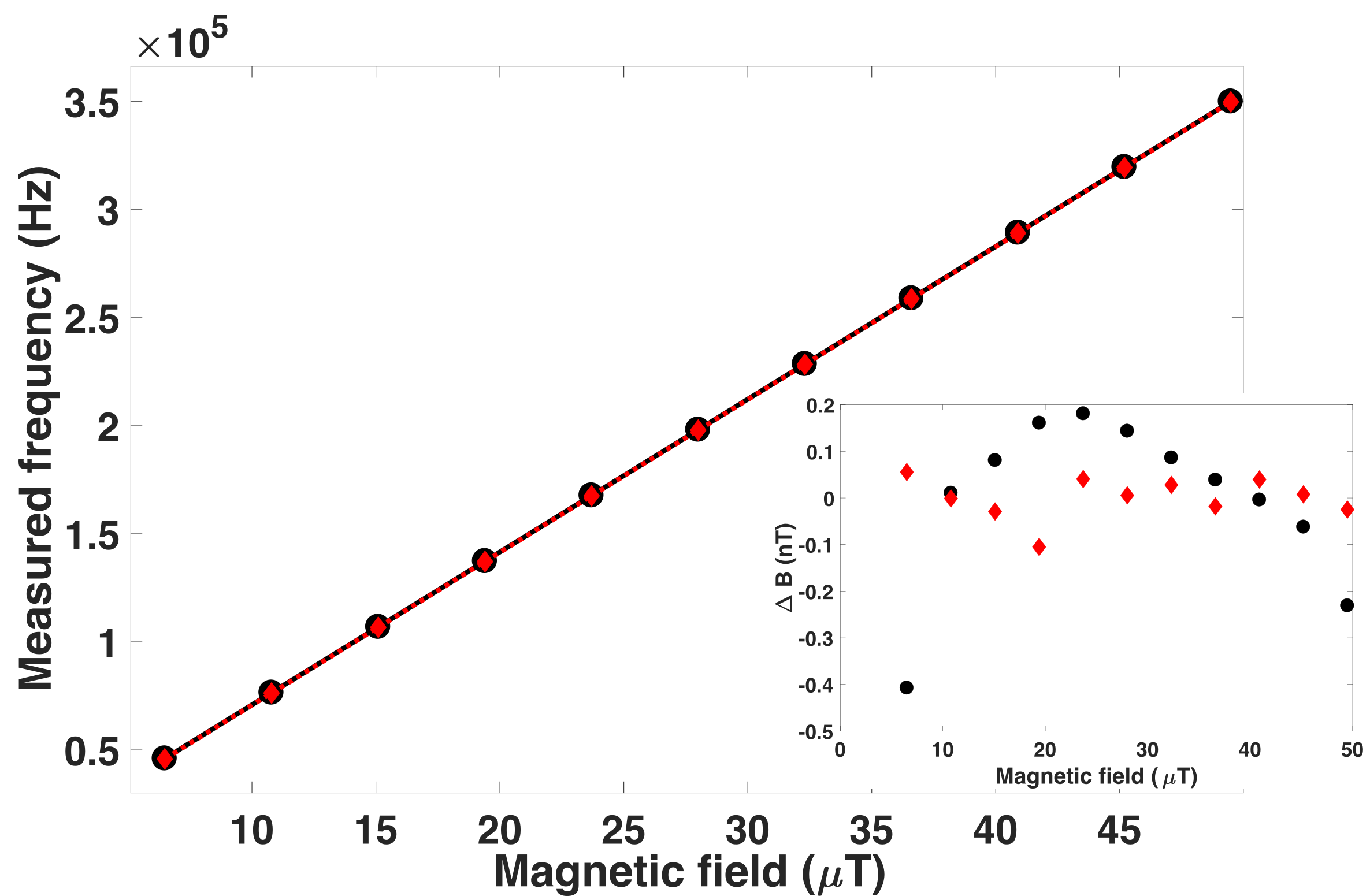}
\caption{The measurement of spin precession frequencies as a function of fields with $P = 0.2$ (black circles) and $P = 0.85$ (red diamonds) with the vertical probe. They are fit to the quadratic polynomial function (dashed lines). The curvature is $C = -16.08 \times 10^{-4} \text{Hz}/\mu \text{T}^2$ at $P = 0.2$ and $C = 2.46 \times 10^{-4} \text{Hz}/\mu \text{T}^2$ at $P = 0.85$. Inset: The residuals of linear fitting of the frequency measurement, expressed in magnetic field unit. The standard deviation is 0.18 nT at $P = 0.2$ (black circles) and 0.045 nT at $P = 0.85$ (red diamonds).}
\label{fig:BvsFMeasurementTot.pdf}
\end{figure}

\begin{figure}
\centering
\includegraphics[width=0.8\columnwidth]{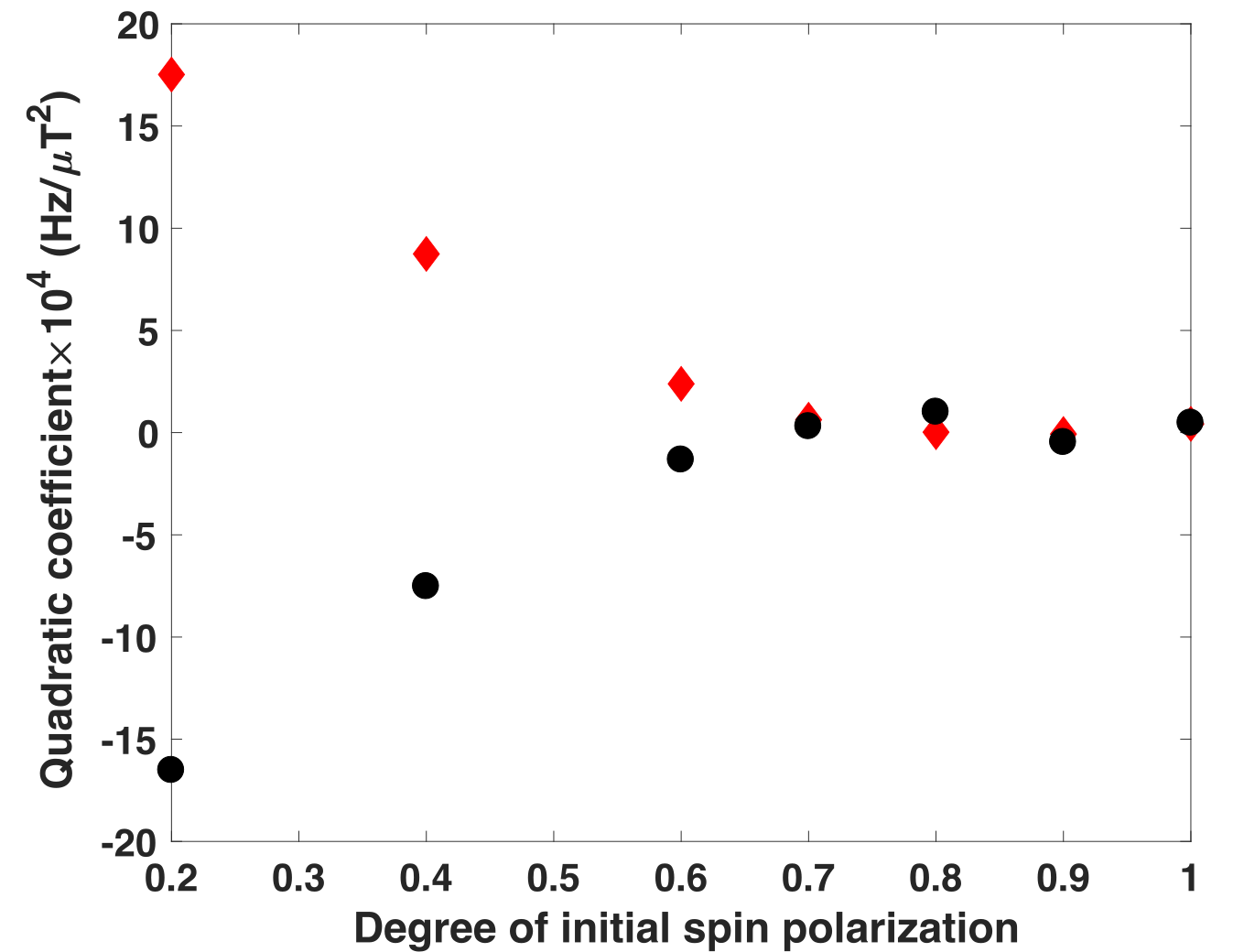}
\caption{The estimation of curvature of spin precession frequency as a function of fields at $\theta = 0^{\circ}$ with different initial spin polarizations. This is based on the numerical simulation of the horizontal probe (red diamonds) and the vertical probe signals (black circles). }
\label{fig:CVsPA90_2.pdf}
\end{figure}

Fig.~\ref{fig:BvsFMeasurementTot.pdf} shows results of averaging the upward and downward field sweep measurements at two different polarizations $P = 0.2$ and 0.85. Even though the first order heading error correction  at $\theta =  0^{\circ}$ is zero from Eq.~\ref{eq:headingerrorcorrectionA}, the low polarization result ($P = 0.2$) shows a curvature of $-16 \times 10^{-4}$ Hz/$\mu$T$^2$, much higher than the curvature for high spin polarization (P=0.85) of $2 \times 10^{-4}$ Hz/$\mu$T$^2$. 

Fig.~\ref{fig:CVsPA90_2.pdf} shows the simulated curvature as a function of polarization. The curvature is negligible in high polarization limit but starts to increase at $P < 0.7$. The curvature $C = -16.52\times 10^{-4}\,\text{Hz}/\mu \text{T}^2$ at $P = 0.2$ with the vertical probe agrees well with experimental measurement. This nonlinear frequency shift is due to the increase in population of $F = 1$ state. The horizontal and vertical probe beams have opposite curvature signs. So averaging of the two signals can cancel the non-linearity of the frequency, similar to the cancellation of the instantaneous frequency oscillations described previously.

 \begin{figure}
\centering
\includegraphics[width=0.95\columnwidth]{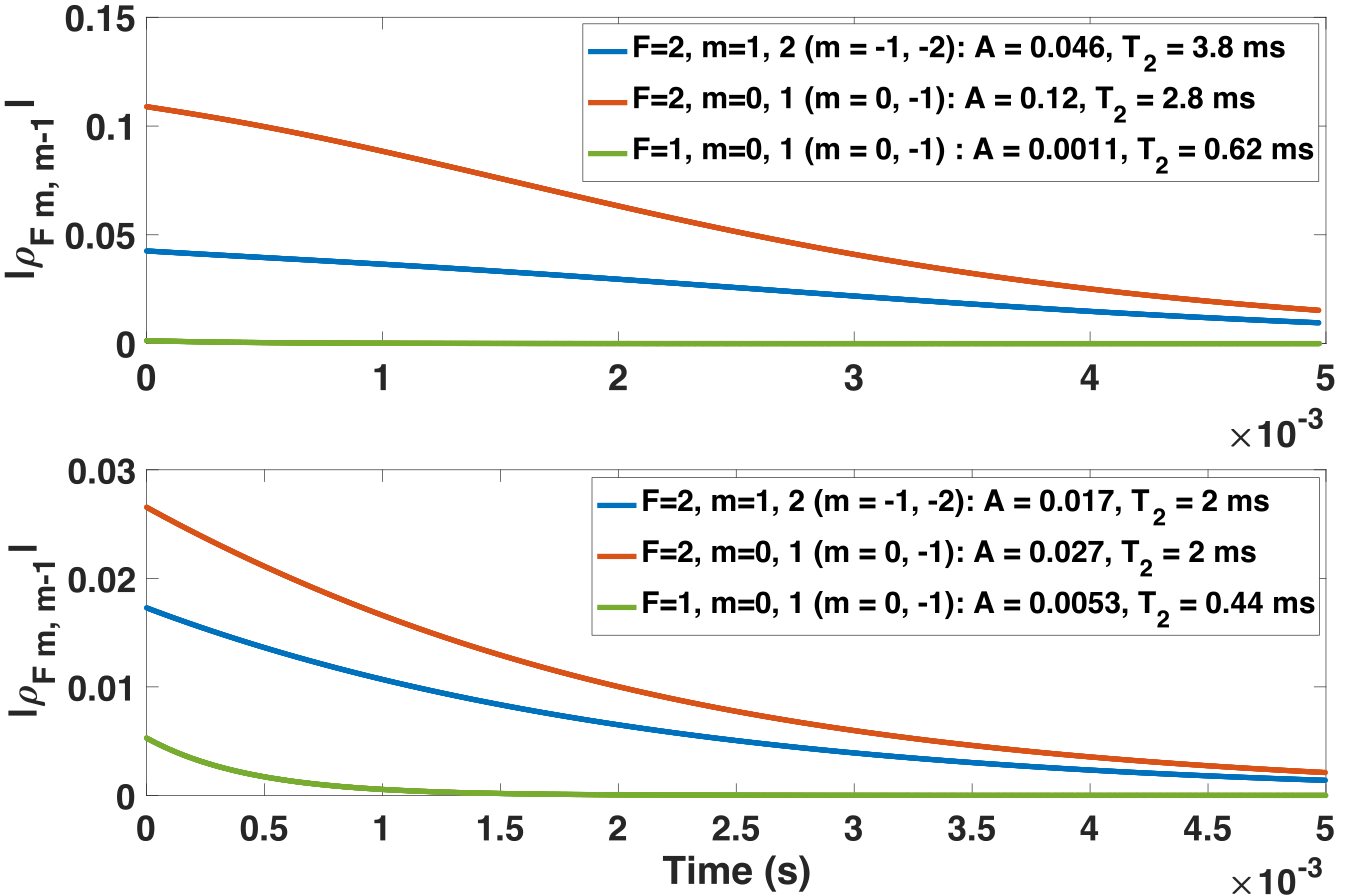}
\caption{Time-evolution of the absolute values of simulated coherences $\rho_{F\,m,m-1}$ at $\theta = 0^{\circ}$ with $P = 0.85$ (top) and $P = 0.2$ (bottom). $F = 2$ state has two distinct coherences (red and blue lines) while $F = 1$ state has one distinct coherence (green line). Each curve is fitted to: $P = Ae^{-t/T_2}$.}
\label{fig:SxCoherences.pdf}
\end{figure}

To understand the origin of the non-linearity we simulate the evolution of individual coherences. When $\theta=0^\circ$ the $F=2$ state has two distinct coherences and $F=1$ state has one distinct coherence. Their initial amplitudes and relaxation are shown in Fig.~\ref{fig:SxCoherences.pdf} for $P=0.85$ and $P=0.2$. The relative strength of $F = 1$ coherence is much higher at $P = 0.2$ than at $P = 0.85$. As the nuclear spin causes splitting of Zeeman frequencies between two hyperfine states, their interference can generate the observed frequency shift at low polarization. To check the origin of the non-linearity we nulled the nuclear magnetic moment in the simulation by setting $g_I = 0$. The curvature at $P = 0.2$ then reduced to $C = 0.2 \times 10^{-4} \,\text{Hz}/\mu \text{T}^2$. This suggests that the nonlinear frequency shift at low polarization is interestingly caused by the linear Zeeman interaction of the nuclear magnetic moment. 

\section{Conclusion}

In this paper we have studied heading errors in $^{87}\text{Rb}$ magnetometer as a function of both the direction and magnitude of magnetic field at different initial spin polarizations. The novel double-probe sensor has shown high sensitivity and significant heading error suppression. 

In the high spin polarization limit, we can correct for heading errors by using analytical expression which is derived based on the density matrix formalism. With the correction, the measured field accuracy is about 0.1 nT in a $50\,\mu \text{T}$ Earth's field, suppressing heading errors by two orders of magnitude. We verify linearity of the measured Zeeman frequency with respect to the field up to Earth's field with a deviation of less than 0.05 nT. At lower polarization, we observe additional heading errors due to the difference in Larmor frequency of the $F = 1$ and $F = 2$ states. This generates beating in the measured frequency, and it is no longer linear with the magnetic field. Numerical simulation shows that this nonlinearity is interestingly caused by the linear Zeeman interaction of the nuclear magnetic moment. To cancel these frequency shifts, we average measurements from two orthogonal probe beams that measure opposite relative phases between the two hyperfine coherences. 

These results are useful in reducing systematics of alkali-metal-vapor atomic magnetometers operating at geomagnetic fields, especially those in navigation systems \cite{Canciani2017,Fu2020,Bevan2018,Shockley2014,Goldenberga}. We suggest methods of cancelling heading errors with wide range of spin polarizations, and the pump-probe geometry presented in this paper can give a real-time correction of heading errors. Furthermore, the use of a small sensor and VCSEL lasers makes it suitable for development of compact and miniaturized sensors \cite{Kitching2018}. 

\begin{acknowledgments}
This work was supported by the DARPA AMBIIENT program.
\end{acknowledgments}

\appendix \label{sec:Appendix}

\section{$^{87}\text{RB}$ GROUND ENERGY LEVELS} \label{sec:Appendix-1}

The ground state Hamiltonian of $^{87}\text{Rb}$ atoms in the presence of external magnetic field is \cite{2010a}
\begin{equation}\label{eq:Hamiltonian}
H = A_\text{hf}\mathbf{I}\cdot\mathbf{S} + g_s\mu_B\mathbf{S}\cdot\mathbf{B} - g_I\mu_B\mathbf{I}\cdot\mathbf{B}
\end{equation}
where $A_\text{hf}$ is the hyperfine constant, $\mathbf{I}$ is the nuclear spin, $\mathbf{S}$ is the electron spin, $\mathbf{B}$ is the field vector, and $g_S$ and $g_I = \mu_I/(\mu_B I)$ are respectively the electron and nuclear g-factors. The first term corresponds to the hyperfine interaction, and the later terms represent the Zeeman interaction of electron and nuclear spin respectively. If we define $\mathbf{F} = \mathbf{I} + \mathbf{S}$ as the total atomic angular momentum, each hyperfine state contains $2F + 1$ magnetic sublevels. The eigenvalue of the Hamiltonian gives the energy of the state, which is described in Eq.~\ref{eq:BreitRabi}. This can be further simplified in Earth's field using $x \ll 1$ as \cite{Seltzer2007} 
\begin{equation}\label{eq:BreitRabi2}
E(F,m) = - \frac{\hbar \omega_\text{hf}}{2(2I+1)} + (\pm \mu_\text{eff} - g_I \mu_B)Bm \mp \frac{\mu^{2}_\text{eff} B^2 m^2}{\hbar \omega_\text{hf}} \end{equation}
where $\mu_\text{eff} = (g_s \mu_B + g_I \mu_B)/ (2I+1)$. 

\begin{figure}
\centering
\includegraphics[width=0.85\columnwidth]{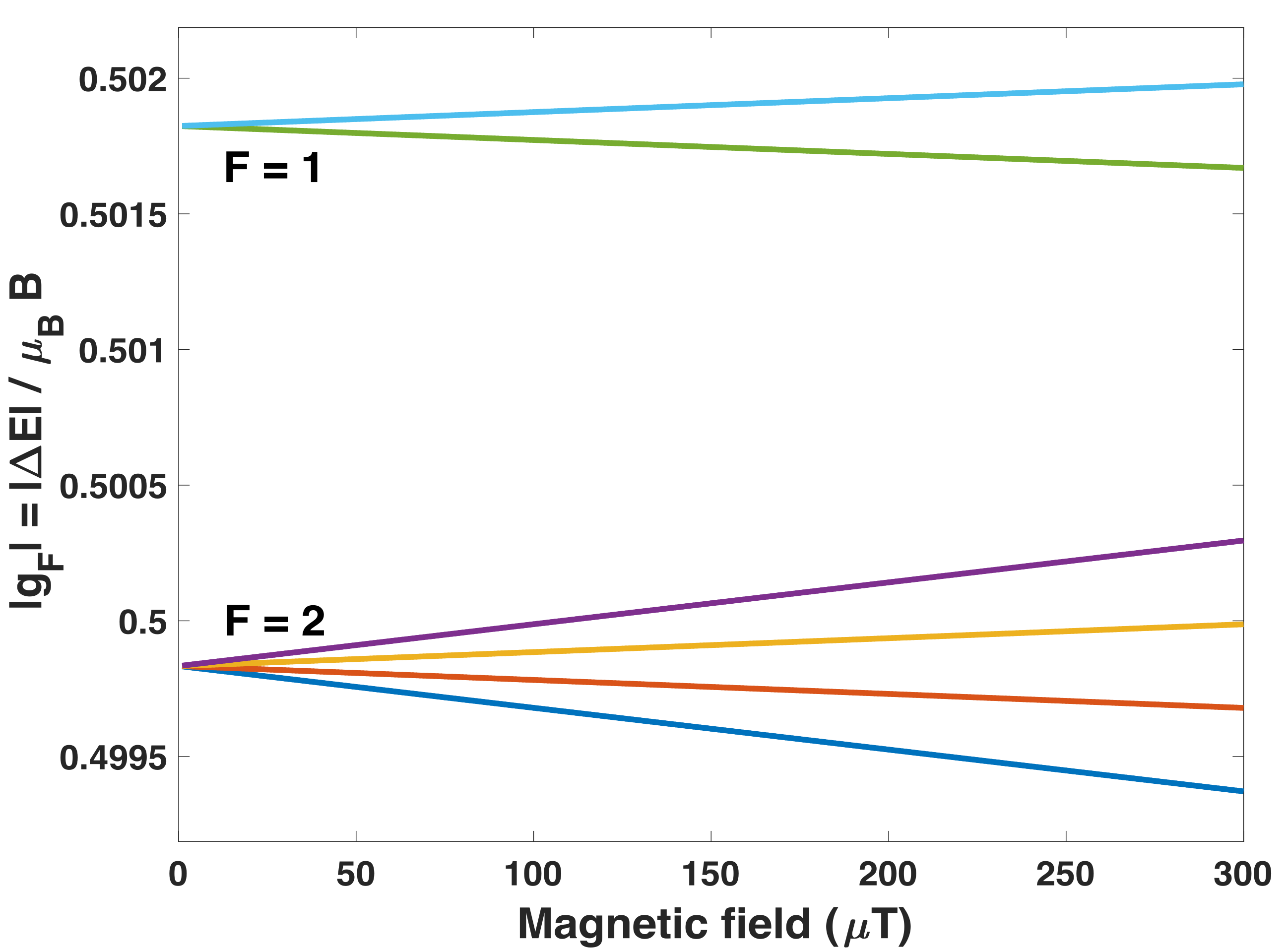}
\caption{The Land$\acute{\text{e}}$ factor $g_F = \Delta E/\mu_B B$ for $m \rightarrow m-1$ Zeeman-transitions for $^{87} \text{Rb}$ atoms in ground state.}
\label{fig:gFactor.png}
\end{figure}

The Land$\acute{\text{e}}$ factors for $m \rightarrow m-1$ Zeeman-transitions from Eq.~\ref{eq:BreitRabi} are shown in Fig.~\ref{fig:gFactor.png}. For $^{87}$Rb atoms in  geomagnetic-strength fields the difference between Zeeman resonances for $F=1$ and $F=2$ states dominates over nonlinear splitting within each hyperfine state.

\section{THE DENSITY MATRIX FORMALISM} \label{sec:Appendix-2}

Instead of a single wave function, we use a density matrix to describe an ensemble of atoms in a mixed state. For $N$-number of atoms, the density operator is 
\begin{equation}\label{eq:densityoperator}
\rho = \frac{1}{N}\sum_i \left| \psi_i \right>\left< \psi_i \right|
\end{equation}
where $\left|\psi\right>$ is the single wave function. The evolution of the ground density matrix of $^{87}\text{Rb}$ atoms is given by \cite{Appelt1998}
\begin{align}\label{eq:DensityEvolution}
\frac{d}{dt}\rho &= \frac{1}{i\hbar}\left[H, \rho \right] + R_\text{SE}\left[\varphi\left(1+ 4\left< \mathbf{S}\right>\cdot\mathbf{S} \right) -\rho \right] \\
&+ R_\text{SD}\left[\varphi - \rho \right] + R_\text{OP}\left[\varphi \left(1+2\mathbf{s}\cdot\mathbf{S}\right) - \rho \right] + D\nabla^2\rho. \nonumber
\end{align}
The first term corresponds to the evolution from the ground state free-atom Hamiltonian of Eq.~\ref{eq:Hamiltonian}. It includes the hyperfine coupling and the Zeeman interaction of the electron and nuclear spin with the external magnetic field. In addition, there is spin relaxation due to several collisional mechanism. The second term describes the spin-exchange collisions between $^{87}\text{Rb}$ atoms where $R_\text{SE}$ is the spin-exchange rate, $\left< \mathbf{S}\right>$ is the expectation value of spin, and $\mathbf{S}$ is the spin operator. This process preserves total spins but redistributes populations in the hyperfine states such that it destroys the spin coherence. This is because the two hyperfine states precess in opposite directions. The third term characterizes the spin-destruction collisions of $^{87}\text{Rb}$ atoms with other $^{87}\text{Rb}$ atoms and buffer gas $\text{N}_2$ molecules where $R_\text{SD}$ is the spin-destruction rate. The spin-destruction collisions destroy the total spin of colliding atoms. The general collisional rate is given by
\begin{equation}
R = n\sigma\bar{v}
\end{equation} 
where $n$ is the atomic number density, $\sigma$ is the effective collisional cross-section, and $\bar{v} = \sqrt{8 k_B T/ \pi M}$ is the relative thermal velocity between the colliding atoms. Here $T$ is the temperature, $k_B$ is the Boltzmann constant, and $M$ is the reduced mass of the atoms. We operate with $^{87}$Rb density on the order of $10^{12}/\text{cm}^3$, where the spin exchange rate between alkali-metal atoms dominates over other relaxation rates. This allows us to measure the $^{87}$Rb density from the transverse spin relaxation time $T_2 = 8/R_\text{SE}$ in the low polarization regime with the spin-exchange cross-section $\sigma_\text{SE} = 1.9\times 10^{-14}\,\text{cm}^2$ \cite{Sheng2013}. The fourth term describes the optical pumping effect where $R_\text{OP}$ is the optical pumping rate. Here $\mathbf{s} = i\mathbf{e}\times\mathbf{e}^{*}$ is the mean photon spin vector where $\mathbf{e} = \mathbf{E}/|\mathbf{E}|$ is the polarization unit vector. The fifth term characterizes diffusion of $^{87}\text{Rb}$ atoms to the cell walls in the presence of buffer gas molecules where $D$ is the diffusion constant. 

The density matrix of an alkali atom can be decomposed into a purely nuclear part which is 
\begin{equation}\label{eq:DensityNuclear}
\varphi = \frac{1}{4}\rho + \mathbf{S}\cdot\rho\mathbf{S}.
\end{equation}
The collisions and optical pumping are sudden with respect to the nuclear polarization such that they only destroy the electron polarization and preserve the nuclear part as shown in Eq.~\ref{eq:DensityEvolution}. 

The transverse spin component is a combination of coherences oscillating at different frequencies. Its expectation value is 
\begin{eqnarray} \label{eq:TransSpin}
& \left< S_x \right> & =  \text{Tr}\left[\rho S_x \right] = \sum_{F\,m' = m \pm 1} A_{F\,m,m'} \rho_{F\,m,m'}  \\ \nonumber
& = &\frac{1}{2(2I+1)} ( \sqrt{(I+1/2-m)(I+3/2+m)} \, \rho_{I+1/2\, m, m+1}  \\ \nonumber
& + &  \sqrt{(I+1/2+m)(I+3/2-m)} \, \rho_{I+1/2\, m, m-1} \\ \nonumber
& - &  \sqrt{(I+1/2+m)(I-1/2-m)} \, \rho_{I-1/2\, m, m+1} \\ \nonumber
& - &  \sqrt{(I+1/2-m)(I-1/2+m)} \, \rho_{I-1/2\, m, m-1} ). \\ \nonumber
\end{eqnarray}
For $\Delta m = \pm 1$ Zeeman transitions, $F = 2$ state has four coherences while $F = 1$ has two coherences. 

If the probe laser is far-detuned from the D1 line of $^{87}\text{Rb}$ atoms such that the detuning is much larger than the ground state hyperfine splitting, its optical rotation angle is proportional to $\left< S_x \right>$ as~\cite{Sheng2013}
\begin{equation}
\phi = \frac{c r_e f_{osc} n l}{\left(\nu-\nu_\alpha\right)} \left<S_x \right>
\end{equation}
where $r_e = 2.82 \times 10^{-13}$ cm is the classical electron radius, $f_{osc} = 0.34$ is the oscillator strength of the D1 transition of $^{87}\text{Rb}$ atoms, $n$ is the vapor density, $l$ is the path length of the probe beam, and $\left( \nu-\nu_\alpha \right)$ is the probe detuning. 

\section{DERIVATION OF AN ANALYTICAL EXPRESSION OF THE HEADING ERROR CORRECTION} \label{sec:Appendix-3}

Due to the nontrivial energy structure of $^{87}\text{Rb}$ atoms, the measured spin precession frequency is a weighted sum of different Zeeman transition frequencies. Let us assume that the spin is initially fully polarized ($P = 1$). The system is in a pure state as all spins are along the magnetic field $B_z$, which is $\rho_0 = \left| 2 \, 2\right>\left< 2 \, 2 \right|$. We then rotate the spin about $+y$ by angle $\theta'$ relative to the field. This is equivalent to applying the rotation operator $D_y = e^{-i\theta' F_y}$ which is represented by the Wigner D-matrix as $d^F_{m',m}(\theta') = \left<F\,m' | e^{-i\theta' F_y} | F\,m\right> = D^F_{m'm}(0,\theta',0)$. The coherence between $m$ and $m - 1$ is given by  
\begin{align}\label{eq:coherence}
\rho_{F =2\, m, m-1 } & = \left<2\, m-1|D^{\dagger}_y\rho_o D_y |2\, m\right> \nonumber  \\
& = \left<2\, m-1|e^{i\theta' F_y}|2\, 2\right>\left<2\, 2|e^{-i\theta' F_y}|2\, m\right> \nonumber \\ 
& = D^{2*}_{2\, m-1}(0,\theta',0)D^{2}_{2\, m}(0,\theta',0) \nonumber \\
& = d^{2*}_{2\, m-1}(\theta')d^2_{2\, m}(\theta')
\end{align}
where $m$ ranges from -1 to +2. By using Eq.~\ref{eq:BreitRabi2}, the weighted sum of the Zeeman transition frequencies up to second order is 
\begin{align}\label{eq:P1Anal}
\omega_{tot} & = \frac{\sum_{m = -1}^{2} A_{2 \,m,m-1} \omega_{2\, m} \rho_{2\, m,m-1}}{\sum_{m = -1}^{2} A_{2 \,m,m-1} \rho_{2\, m,m-1}} \nonumber \\
& = \frac{1}{4\hbar}(-3g_I+g_S)\mu_B B - \frac{3 (g_I+g_S)^2 \mu_B^2 \text{sin}\theta B^2}{32 A_\text{hf} \hbar}
\end{align} where $A_{2 \,m,m-1}$ is derived from Eq.~\ref{eq:TransSpin}. Here $\theta = 90^\circ - \theta'$ is the angle between the spin orientation and the nominal magnetometer orientation where the field is perpendicular to the spin as shown in Fig.~\ref{fig:MagnetometerBasics.pdf}.

The pulsed $^{87}\text{Rb}$ magnetometer can achieve a high initial spin polarization, about $P = 0.9$ where $F = 1$ state is almost depopulated. In this condition the atom is in a mixed state of $F = 2$ sublevels which approximately follows the spin-temperature distribution \cite{Anderson1960}. For high optical pumping rate for a pressure broadened optical line, one also reaches spin-exchange equilibrium \cite{Appelt1998}. This is analogous to the thermal equilibrium, and the relative population in each Zeeman sublevel is described by the spin-temperature distribution: 
\begin{equation}
\rho\,(F,m) = \frac{e^{\beta m}}{Z}
\end{equation} where $Z = \sum_{m = -F}^{F} e^{\beta m}$ is the partition function and $\beta = \text{ln}[(1+P)/(1-P)]$ is the spin temperature that depends on the spin polarization $P = 2\left< S_x \right>$. Therefore the initial atomic state is $\rho_0 = \sum_{m = -2}^{2} \rho_{2\,m} \left| 2 \, m\right>\left< 2\, m \right| = \sum_{m} \left( e^{\beta m}/Z \right) \left| 2 \, m\right>\left< 2\, m \right| $. If we apply the rotation operator, each coherence term becomes 
\begin{align}
& \rho_{F = 2\, m, m-1}  = \left<2\, m-1| D^{\dagger}_y\rho_o D_y |2\, m\right>  \nonumber \\
& =  \sum_{m' = -2}^2 \rho_{2\,m'}\left<2\, m-1|e^{i\theta' F_y}|2\, m'\right>\left<2\, m'|e^{-i\theta' F_y}|2\, m\right> \nonumber \\ 
& = \sum_{m' = -2}^{2} \frac{e^{\beta m'}}{Z} {d^{2*}_{m' m-1}(\theta')}d^2_{m' m}(\theta'). 
\end{align} where $m$ ranges from -1 to +2. The weighted sum of the Zeeman transition frequencies is then 
\begin{align}\label{eq:PL1Anal}
\omega_{tot} & = \frac{\sum_{m = -1}^{2} A_{2\,m,m-1} \omega_{2\, m} \rho_{2\, m,m-1}}{\sum_{m = -1}^{2} A_{2 \,m,m-1} \rho_{2\, m,m-1}} \nonumber \\
& = \frac{1}{4\hbar}(-3g_I+g_S)\mu_B B \nonumber \\
& - \frac{3(g_I+g_S)^2 \mu_B^2 \text{sin}\theta B^2}{32 A_\text{hf} \hbar}\frac{P(7+P^2)}{5+3P^2}.
\end{align} It has an additional polarization dependent factor and reduces to Eq.~\ref{eq:P1Anal} at full polarization ($P = 1$). 

Inverting the equation above results in 
\begin{equation}\label{eq:headingerrorcorrection}
B = \frac{4h\nu}{(g_S - 3g_I)\mu_B}\left[1 - \frac{3\nu}{\nu_\text{hf}}\text{sin}\theta \frac{P(7+P^2)}{5+3P^2}\right]
\end{equation}
where $\nu = \omega/2\pi$ is the measured Larmor precession frequency. To simplify this equation we set $g_I$ to zero in the second term. 

\section{NUMERICAL SIMULATION METHOD}\label{sec:Appendix-4}

We simulate the optical rotation signal by solving the density matrix equation of Eq.~\ref{eq:DensityEvolution} with parameters matching the experimental condition and calculate $\left< S_x \right>$ with Eq.~\ref{eq:TransSpin}. The spin-exchange rate is $R_\text{SE} = n_\text{Rb}\sigma_\text{SE}\bar{v}_\text{RbRb} = 3.6 \times 10^3 \;1/\text{s}$ where $n_\text{Rb} = 4.4\times 10^{12}/\text{cm}^3$, $\sigma_\text{SE} = 1.9\times 10^{-14}\,\text{cm}^2$, and $\bar{v}_\text{RbRb} = 430 \,\text{m}/\text{s}$. The relative thermal velocity is calculated at cell temperature $T = 100\,^{\circ}\text{C}$. The spin-destruction rate is $R_\text{SD} = n_\text{Rb}\sigma_{\text{Rb}\text{Rb}}\bar{v}_\text{RbRb} + n_{\text{N}_2}\sigma_{\text{N}_2\text{Rb}}\bar{v}_{\text{Rb}\text{N}_2} = 110 \;1/\text{s}$ where $\sigma_{\text{Rb}\text{Rb}} = 9\times 10^{-18} \,\text{cm}^2$, $\sigma_{\text{Rb}\text{N}_2} = 1\times 10^{-22}\, \text{cm}^2$ \cite{Allred2002}, $\bar{v}_{\text{Rb}\text{N}_2} = 620 \,\text{m}/\text{s}$, and $n_{\text{N}_2} = 1.7 \times 10^{19}/\text{cm}^3$. We fit the simulated signal to Eq.~\ref{eq:ExpDecaySine} for $T_m = 3$ ms to estimate the spin precession frequency. The simulated signal has a slightly longer $T_2$ than the measured signal by 0.5 ms. This can be due to field gradients or diffusion broadening \cite{Lucivero2017}.

\bibliographystyle{apsrev4-1}
\bibliography{HeadingErrors_Arxiv}

\end{document}